\documentstyle[11pt,mypp,epsf]{article}

\def\deg{$^{\circ}$}

\def\arp{Arp~102B}
\def\mcg{\mbox{MCG-6-30-15}}
\def\rg{$R_g$}

\def\gsim{\lower 2pt \hbox{$\, \buildrel {\scriptstyle >}\over {\scriptstyle
\sim}\,$}}
\def\lsim{\lower 2pt \hbox{$\, \buildrel {\scriptstyle <}\over {\scriptstyle
\sim}\,$}}

\newcommand{\beq}{\begin{equation}}
\newcommand{\eeq}{\end{equation}}
\newcommand{\comment}[1]{}

\begin{document}

\markboth{Bromley, Chen \& Miller}{Rotating Black Holes}
\pagestyle{myheadings}
\thispagestyle{empty}

\vspace{33pt}

\begin{center}
{\Large \bf
Line Emission from an Accretion Disk around a Rotating \\ 
\vspace{2pt}
Black Hole: Toward a Measurement of Frame Dragging
}

\vspace{10pt}

{\normalsize \sc  Benjamin C. Bromley$^{1,2}$, 
   Kaiyou Chen$^{1,3}$ and Warner A. Miller$^1$}

\vspace{10pt}

\parbox[t]{5.75in}{
 {\footnotesize \mbox{}$^1$Theoretical Astrophysics, MS B288, 
  Los Alamos National Laboratory, Los Alamos, NM 87545} \\
 {\footnotesize \mbox{}$^2$Theoretical Astrophysics, 
  MS-51, Harvard-Smithsonian
  Center for Astrophysics   \vspace{-3pt}
  \\
  \mbox{} \, 60 Garden Street, Cambridge, MA 02138} \\
 {\footnotesize \mbox{}$^3$Columbia Astrophysics Laboratory, 
  Columbia University, 538 West 120th Street,
  New York, NY 10027
 }
}
\end{center}

\begin{abstract}

     Line emission from an accretion disk and a corotating hot spot
about a rotating black hole are considered for possible signatures of
the frame-dragging effect. We explicitly compare integrated line
profiles from a geometrically thin disk about a Schwarzschild and an
extreme Kerr black hole, and show that the line profile differences
are small if the inner radius of the disk is near or above the
Schwarzschild stable-orbit limit of radius $6GM/c^2$. However, if the
inner disk radius extends below this limit, as is possible in the
extreme Kerr spacetime, then differences can become significant,
especially if the disk emissivity is stronger near the inner regions.
We demonstrate that the first three moments of a line profile define a
three-dimensional space in which the presence of material at small
radii becomes quantitatively evident in broad classes of disk models.
In the context of the simple, thin disk paradigm, this moment-mapping
scheme suggests formally that the iron line detected by the Advanced
Satellite for Cosmology and Astrophysics mission from MCG-6-30-15
(Tanaka et al. 1995) is 3 times more likely to originate from a disk about
a rotating black hole than from a Schwarzschild system. A
statistically significant detection of black hole rotation in this way
may be achieved after only modest improvements in the quality of data.
We also consider light curves and frequency shifts in line emission as
a function of time for corotating hot spots in extreme Kerr and
Schwarzschild geometries. The frequency-shift profile is a valuable
measure of orbital parameters and might possibly be used to detect
frame dragging even at radii approaching $6GM/c^2$ if the inclination
angle of the orbital plane is large. The light curve from a hot spot
shows differences as well, although these too are pronounced only at
large inclination angles.

\vspace{5pt}

{{\it Subject headings:} }
 accretion, accretion disks --- black holes --- galaxies: active ---
 galaxies: nuclei --- line profiles

\end{abstract}

\section{Introduction}

Black holes manifest themselves, in part, through emission of intense
radiation from material associated with accretion processes.
Currently, about a dozen black hole candidates are known within the
Galaxy and many more supermassive black holes are believed to be in
the nuclei of active galaxies. It is likely that most, if not all, of
these black holes are spinning near their maximal rate such that
the angular momentum $J$ and mass $M$ are related by $J \sim G
M^2/c$. Thus light emitted from the region close to the horizon of a
rotating black hole is subject to frame dragging --- an effect
predicted by Einstein's theory of general relativity, also known as
gravitomagnetism or the Lense-Thirring effect (Thirring \& Lense 1918;
Misner \& Thorne \& Wheeler 1970).

To date frame dragging has not been verified by any observation or
measurement.  However, there are currently two proposed satellite
experiments, LAGEOS-3 and Gravity Probe-B, which are slated to measure
this effect around the earth (Ciufolini 1986; Habib et.~al. 1994; Van
Patten \& Everitt 1976).  While the frame dragging on satellite orbits
around the earth is small (approximately 1 part in $10^7$) when
compared with the classical quadrupole contributions, in astrophysics
the gravitational binding energies are large enough so as to make the
dragging of inertial frames the dominant precessional effect.
Furthermore, the frame dragging effect together with the shear
viscosity of accreted material may be responsible for the alignment of
an accretion disk with the parent black hole's spin axis, and may
explain the stability of astrophysical jets (Bardeen \& Peterson
1975).  It is for these reasons that we search for an astrophysically
relevant setting to observe frame dragging.

Material drawn into the neighborhood of a black hole is likely to
carry some net angular momentum, hence the formation of an accretion
disk seems highly probable if not inevitable.  The dominant radiation
from such an object is in the form of continuum emission, which has
been the subject of intensive study over last few decades.  Models of
thermal emission from a geometrically thin and optically thick disk
give good fits to the UV, optical, and IR spectra of many active
galactic nuclei (Sun \& Malkan 1988).  However, kinematic information
about the disk is difficult to extract from continuum emission. For
example, even with the simplest accretion disk model the
interpretation of continuum spectra is not unique (Ferland \& Rees
1988).  Therefore continuum emission provides only limited
information about the physical environment near the black hole.

Atomic line emission can also arise from accretion disks around supermassive
black holes in AGNs, as evidence accumulated over the last seven years
demonstrates:

\noindent 
1. Observed broad and double-peaked low-ionization lines from some
AGNs are consistent with emission from a geometrically thin Keplerian
disk. \arp\ (Chen, Halpern \& Filippenko 1989; Chen \& Halpern 1989)
is the prototype and more samples are listed in Eracleous \& Halpern
(1994).  Nearly all of these disk emitters are found within radio
galaxies.

\noindent 
2. In contrast to low-ionization lines, high-ionization
lines such as Ly$_\alpha$ of \arp\ lack the corresponding double peaked
components. The material producing the low-ionization double-peaked
line emission must be denser than typical broad-line clouds. This
requirement is in agreement with an accretion disk model (Halpern et
al. 1995).

\noindent 
3. The double peaked line emission of \arp\ is polarized and the
polarization angle is constant (Antonucci, Agol \& Hurd 1996) for the
whole line-emission profile. This requires a flat, disk-like
geometry for scattering electrons.

\noindent 
4. The physical parameters determined from fitting the detailed line
profile with the accretion disk model is consistent with those
inferred from other independent methods.  Inclination angles deduced
from a relativistic disk model are generally modest ($\sim 30^\circ$),
consistent with fact that many of the objects are double-lobe radio
sources in which broad emission lines are neither obscured by the host
galaxy nor emitted from material with relativistic line-of-sight
velocity.  In particular, for radio galaxy 3C390.3, the inclination
angle of $\sim 28^\circ$ inferred from the observed superluminal
motion is extremely close to the $26^\circ$ value derived from the fit
with the relativistic disk model (Eracleous, Halpern \& Livio 1996).
 
Since the physics of atomic line emission is well understood, the
information carried by line emission provides us with a useful tool to
explore the region near the black hole.  At present, optical lines
provide the best evidence of black holes in the nuclei of active
galaxies.  However, the typical orbit radius of the disk region
emitting optical lines is a few hundred times the gravitational radius
$\mbox{\rg} \equiv G M/c^2$ (this is half of the horizon radius in the
case of a nonrotating black hole), not close enough to the black hole
to reveal the higher-order frame dragging effects.  It was speculated
that some X-ray emission lines may come from a region much closer to
the black hole, thus permitting a test of the frame dragging effect
(Chen et al. 1989).  Recent observations by the Japanese ASCA
satellite have turned this speculation into reality.  Broad iron K
lines have been reported for three Seyfert galaxies (Fabian et
al. 1994).  A four-day exposure of one of them, MCG 6-30-15, has
yielded a well-resolved Fe K$_\alpha$ line with an extremely broad
profile that is consistent with an accretion disk extending from 6 -
20 \rg\ (Tanaka et al. 1995). This significant discovery is good cause
to believe that frame dragging can be detected in an astrophysical
setting.

In the present work we focus on the information that can be extracted
from line profiles and time-varying signals from an accretion disk.
Thus, we are embarking on a study in which the detection of frame
dragging is part of a more general probe of the physics in the
environment of a rotating black hole.  This paper reports our
investigation of frame dragging and its effect on the line emission
from an accretion disk about a rotating (Kerr) black hole.  We discuss
the construction of an accretion disk image from photons traveling in
the curved Kerr spacetime geometry, and then consider signatures of
frame dragging from a radiating disk and from a hotspot of localized
emission. This material is a synthesis and extension of the work by
Laor (1991), Karas, Vokrouhlick\'{y} \& Polnarev (1992), and Zakharov
(1994).  The principle advances reported here stem from an increase in
the temporal and photon-frequency resolution in the numerical code
which was brought to bear on the problem of observing the frame
dragging effect.

\section{Ray Tracing in a Kerr Metric and Imaging an Accretion Disk}  

The trajectory of a photon in the Kerr metric can be described
with aid of three constants of motion (Misner, Thorne, \& Wheeler 1970), 
\beq
 \begin{array}{c} 
    E = -p_t \ , \ \ \ \ \ \ L = -p_\phi \ , \\ 
    Q = p^2_\theta - a^2 E^2 \cos^2 \theta + L^2 \cot^2 \theta \ , 
 \end{array}
\eeq
where $r$, $\theta$, $\phi$ and $t$ are the usual Boyer--Lindquist
coordinates, $p$ is the four-momentum and $a=J/M$ is the magnitude of
reduced angular momentum in units the black hole mass (a value of $M$
is the maximal, or extreme, case).  In this section and hereafter, we
use natural units such that $G = c = 1$. In order to eliminate
difficulties with inflection points along any trajectory we solved the
three second-order geodesic equations, for $r(t)$, $\theta(t)$ and
$\phi(t)$.
\beq \label{eq:geodesic}
  {d^2x^i\over dt^2} = -\Gamma^i_{\mu\nu}{dx^\mu\over
      dt}{dx^\nu\over dt} +\Gamma^0_{\mu\nu} {dx^\mu\over dt}{dx^\nu\over
      dt}{dx^i\over dt} \ 
\eeq 
Here, the $\Gamma$'s are connection coefficients; Latin indices take
the values 1,2,3 (with $x^1$, $x^2$ and $x^3$ representing $r$,
$\theta$ and $\phi$, respectively) and Greek indices range from 1 to
4; repeated Greek indices denotes summation.

To obtain an image of an accretion disk as seen by a distant observer
we integrated the geodesic equations (\ref{eq:geodesic}) to obtain
trajectories of photons which propagate from a uniform Cartesian grid
of points in the sky plane to the accretion disk. The uniform grid may
be associated with the field imaged by pixels of a CCD detector.

The integration of equations~(\ref{eq:geodesic}) was performed in
three stages. First, we determined the initial-value data, i.e., the
phase-space location of the photons; the starting points of the
trajectories were typically at a large distance, $10^3 M$, from the
black hole.  In the second stage, we integrated the equations from
$t=0$ until the photon just crossed the plane of the disk. An adaptive
step $4^{\rm th}$-order Runge--Kutta method was used to calculate the
trajectory.  Finally, we used a golden selection search to determine
the spacetime intersection of the photon and the disk with high
accuracy. We discarded photons that were intercepted by the hole
(as determined by an inner cutoff radius) or otherwise missed the
disk. On output we saved the Boyer-Lindquist coordinate of the
photon-disk intersection, the time of flight in the frame of the
distant observer, and frequency shift information for each pixel.

The process of generating a pixel image of an accretion disk requires
tracking $n \times n$ independent photon trajectories. Thus, the
problem is readily implemented on a parallel computer, with each node
assigned to a unique set of pixels. The only pitfall is that some
photon trajectories are more difficult to calculate than others, hence
load balancing is a potential concern. Typically, trajectories which
bring the photon closer to the black hole are more time consuming.
Our solution to the problem was to assign processors to interleaved
rows of the pixel array; the result was a tolerable clocktime
difference of 10--20 \% between the fastest processors and the
slowest. As a result, we were able to generate an image of $1200
\times 1200$ pixels in about 15 minutes using 128 nodes on a Cray T3D.

An example of this procedure can be seen in Figure~1, an image of a
geometrically thin accretion disk around a maximally-rotating black
hole ($a=M$). The inner and outer radii are at $1.25 R_g$ and $10
R_g$, respectively, and the disk is viewed at 75\deg\ with respect
to the polar axis. The false colors indicate frequency shift of
emitted photons as a result of Doppler and gravitational redshift.
The image clearly shows the gravitational refocusing of the photons
emitted from the far side of the disk; this effect causes the far side
of the disk to appear as if it were bent toward the observer. The more
subtle effects of frame dragging are evident in the overall shape of
disk and the asymmetry of frequency shift along contours of fixed
radius between near and far sides of the disk.

\begin{figure}[t]
 \centerline{ \epsfxsize=5in\epsfbox{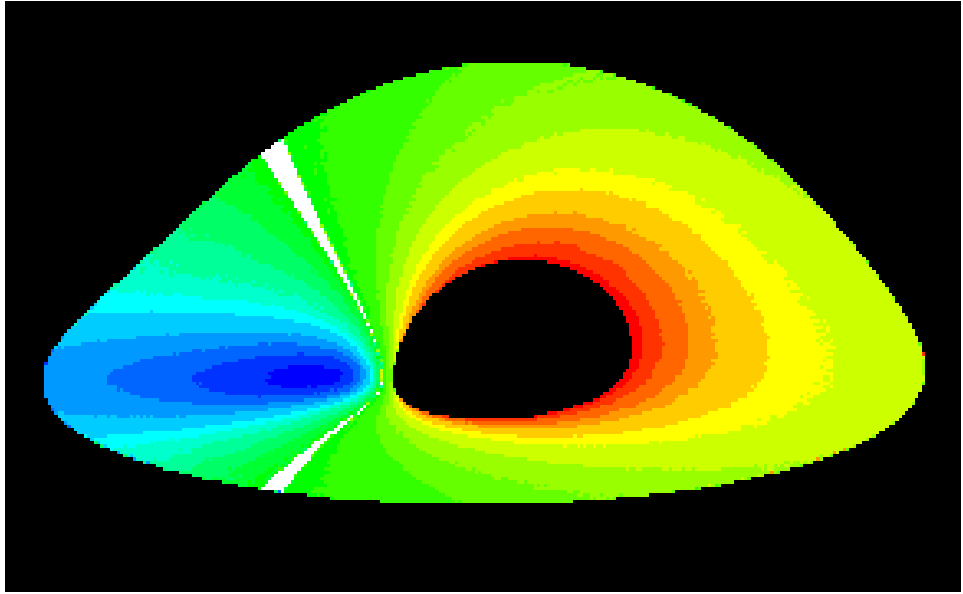} } \caption{ The image
of a geometrically thin and optically thick accretion disk around a
maximally-rotating black hole of mass $M$.  The inner radius and outer
radius are at $1.25 M$ and $10 M$, respectively.  This accretion disk
is viewed at a $75^\circ$ inclination.  The false color contour map
indicates the ratio of observed frequency to emitted frequency, where
the blue shaded bands represent blue shift and red indicates a red
shift.  The white (saturated) patch superimposed on the map contains
the zero frequency shift contour.  [A full color image can be found at
\protect{\tt http://cfata2.harvard.edu/bromley/nu\_nofun.html}.]
}
\end{figure}

\section{Line Profiles of Disk Emission} 

As mentioned above, X-ray line profiles from emission integrated over
the surface of an accretion disk can provide insight into the physics
of the environment immediately surrounding a black hole.  With some
generality we may write the observed flux at frequency $\nu$ from an
accretion disk as
\beq\label{eq:flux}
 F(\nu) = \int \! d\Omega \, \int d\nu_e T(\nu,\mu,d;\nu_e,\mu_e,r_e)
            I(\nu_e,\mu_e,r_e)\, ({\nu \over \nu_e})^3 \ ,
\eeq
where $I$ is the specific intensity which is a function of the
frequency $\nu_e$ of the emitted photon, the angle cosine $\mu_e$ of
the photon emission with respect to the normal of the disk as measured
in the source frame, and the radial location $r_e$ of the emitting
material on the disk. The transfer function $T$ determines the
fraction of locally emitted energy that is ultimately detected by the
observer at frequency $\nu$. In addition to the intrinsic disk
quantities $\nu_e$, $\mu_e$, and $r_e$, this function depends on
observed frequency, the inclination angle of the disk as given by the
angle cosine $\mu$, and the distance $d$ between source and
observer. The integration over $d\Omega$, an element of solid angle,
covers the image of the disk in the observer's sky plane.  The factor
$(\nu/\nu_e)^3$ arises because $I_{\nu}/{\nu^3}=I_{\nu_e}/{{\nu_e}^3}$
is an invariant from one observer to another, and from one event to
another, along the entire trajectory of the photon. This invariant
results from the principle of conservation of photons in a flux tube
together with conservation of volume in phase space (Liouville's
theorem holds in curved spacetime). Indeed, $I_{\nu}/{\nu^3}$ is
equal, up to a power of Planck's constant, to the photon phase-space
density.

In the work of Laor (1991) and Speith, Riffert \& Ruder (1995),
explicit calculation of the transfer function $T$ in
equation~(\ref{eq:flux}) was performed to obtain profiles of
integrated flux versus frequency for a given specific intensity.
Here, we perform essentially the same calculation by binning the
observed flux at each pixel in a disk image.  Figure~2 illustrates the
flux in models where a single atomic line is locally emitted from a
disk with uniform emissivity, so that $I$ is a delta function when
$r_e$ is between the inner and outer disk radii.  The figure shows
line profiles which are similar to those given by Laor (1991), except
that here the resolution is significantly enhanced.

\begin{figure}[t]
 \centerline{
  \epsfxsize=4.0in\epsfbox{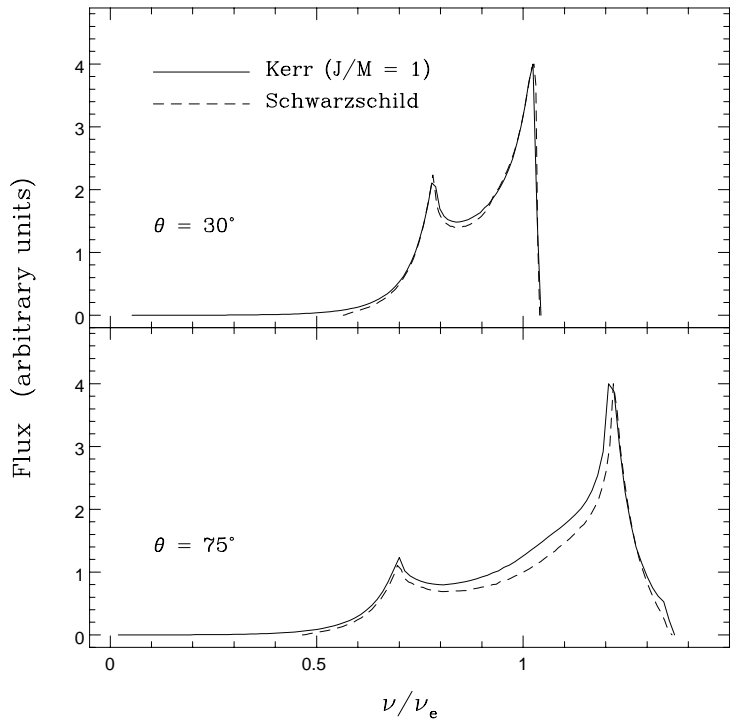}
 }
\caption{
The integrated line flux $F$ as a function of observed frequency $\nu$
(in units of the emitted frequency) for accretion disks in Kerr and
Schwarzschild systems at two different inclination angles, $\theta =
30^\circ$, and $75^\circ$. The inner radius of the disk is 1.25\rg\ in
the Kerr case and 6\rg\ in the Schwarzschild case.  The outer disk
radius is 10\rg\ for both systems. The emissivity is constant over the
disks.
}
\end{figure}

The purpose of Figure 2 is to compare disk emission from an extreme
Kerr black hole with that from a nonrotating (Schwarzschild) system.
The outer disk radius in both systems is 15\rg\ while the inner radius
is 1.25\rg\ and 6\rg\ for the Kerr and Schwarzschild cases
respectively, values which are near or at the innermost stable orbit.
The most notable features in the line profiles are the double peaks
which are mostly determined by the physical and geometrical parameters
at a relatively large distance from the inner most stable orbit.
Thus, the differences between the Kerr and Schwarzschild cases are
expected to be small.  However, the shape of the profiles are
generally distinct. For example, the weak red emission below the red
peak in the line profile is less in the Schwarzschild system.  The
depth of the trough between the two peaks is also less in the
Schwarzschild case than in the extreme Kerr system; at an inclination
angle of 30\deg\ the line flux at the trough minimum is $\sim 5$\%
higher in the Kerr case; at 75\deg\ the excess in the Kerr signal is
about 20\% when the observed frequency equals the emitted frequency.
The effect becomes stronger when the outer disk radius is reduced. For
example, when the outer radius is $r_o = 10$\rg\ the excess in the
Kerr signal is approximately twice that of the $r_o = 15$\rg\ case.

These distinguishing features of the line profiles in the Kerr system
result from material in stable orbits below $\sim 6$\rg. When the disk
emissivity is uniform, this material contributes only a small fraction
of the total flux in a line profile. Thus it is not surprising that
the differences in the line profiles are not pronounced. The detection
of frame dragging by the presence of this material will be difficult,
requiring flux measurement errors to be less than a few percent in
frequency bins that are $\lsim 0.04$ times the width of the line.

Fortunately, the case of uniform disk emissivity is likely to be a
worst-case scenario. Instead, the observational evidence suggests that
the emissivity falls off with radius as a power law of index $\alpha$
such that $0 \lsim \alpha \lsim 4$ (Chen et al., 1989; Eracleus \&
Halpern 1994; Tanaka et al. 1995).  The effect is to brighten the
inner part of the disk relative to the outer part, hence contributions
from emission at small radii are likely to be much greater than in
Figure~2. Indeed, with disk parameters similar to ones fit by Tanaka
et al. (1995) for the \mcg\ Fe line profile we found that the
differences between the Kerr and Schwarzschild cases increased
significantly, as seen in Figure~3.  The figure illustrates that the
height of the central trough and the red peak in an integrated line
profile depends sensitively on material nearest the disk's inner
radius.  The figure also shows that the presence of frame dragging is
not apparent if the inner radius of the disk is $\sim 5 \mbox{\rg}$ or
above; for example the profile from a Kerr system with an inner radius
of $r_i = 5.5$\rg\ is similar to the Schwarzschild case with disk
material down to its minimum stable orbit of 6\rg.  This result gives
hope that black hole rotation may be detected but only if there is a
line-emitting accretion disk that extends well below 6\rg.

\begin{figure}[tbhp]
 \centerline{
  \epsfxsize=3.5in\epsfbox{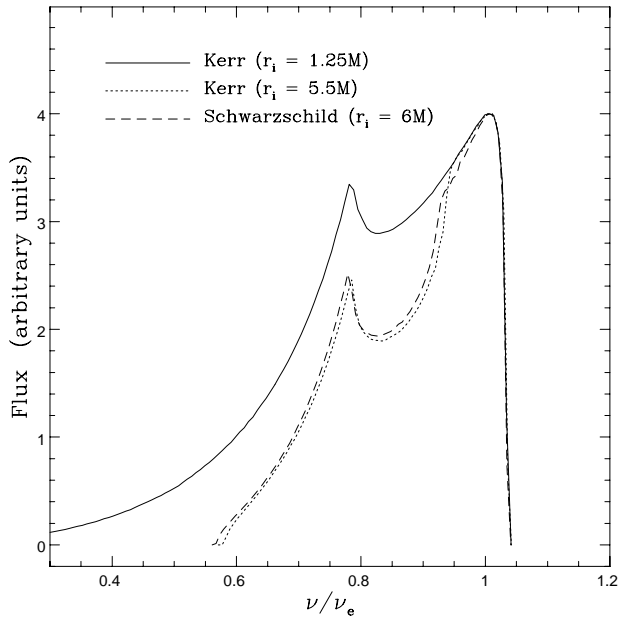}
 }
\caption{
The integrated line flux $F$ from accretion disks, similar to Figure
2.  Here, the disk parameters were chosen to be in the range indicated
by the Tanaka et al. (1995) analysis of \mcg: the emissivity function
of the disk varies as the inverse square of the radius; the
inclination angle is 30\deg; and the outer radii are set to 15\rg.
Note that the Kerr system with inner radius of 5.5\rg\ closely mimics
the Schwarzschild case with inner radius 6\rg.
}
\end{figure}

In practice, detection of a disk with a small inner radius will
require the ability to locate the peaks of the line profile to within a
few percent and to determine flux to similar accuracy.  The next
generation of X-ray satellites beginning with the planned Advanced
X-ray Astronomical Facility could provide the necessary resolution in
energy and flux.  The emissivity of the disk must then be modeled and
fit along with inner and outer disk radii and angle of
inclination. There are prominent features in the line profiles of
Figure~3 that offer, at the very least, consistency checks on a
model. These include the height and location of the peaks, the depth
of the trough, and the kink found on the blue side of the trough when
the inner radius is above several \rg.  Furthermore, some parameters
of the disk model may be determined from other astronomical
observations. For example, the inclination angle of the disk may be
inferred from the apparent superluminous motion of jets or from the
equivalent width of emission lines.

A comparison between a model line profile and currently available
data is shown in Figure~4. The data are from \mcg\ as given in Tanaka
et al. (1995).  The model for this simple demonstration came from a
nonlinear simplex search fit to the disk radii and equivalent width.
A more general fit which includes the inclination angle as a
parameter, requires a series of disk images, each at a distinct
inclination angle.

\begin{figure}[tbhp]
 \centerline{
  \epsfxsize=3.5in\epsfbox{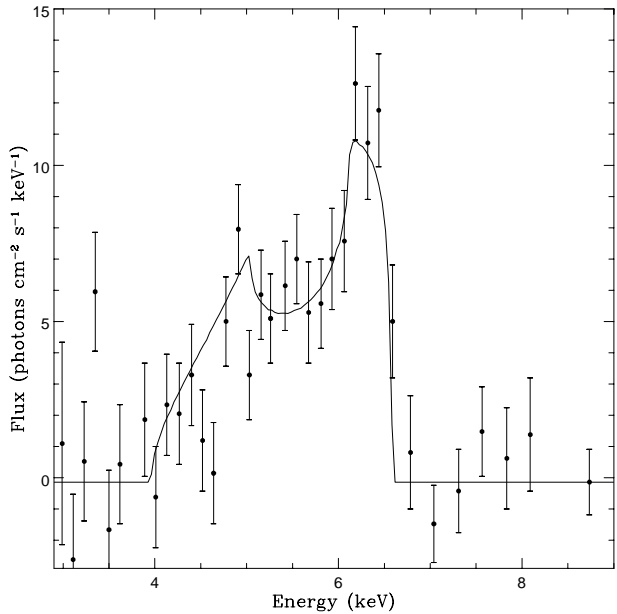}
 }
\caption{
A model superimposed on the Tanaka et al. (1995) data from \mcg. The solid
curve shows a disk with an inner radius of 7.2\rg, an outer radius of 
15.8\rg, and an emissivity index of $\alpha = 3$. The inclination angle
is 30\deg.
}
\end{figure}

\subsection{Line Profile Moments}

The detection of frame dragging should come from fitting models to
data in order to identify disk material in stable orbits below 6\rg.
However, it is reasonable to seek general characteristics of a line
profile that are sensitive to the presence of emitting material at
small radii even if these characteristics lack sufficient information
to uniquely specify a disk model.  Here we look at low-order moments
of the line profile; other possible characteristics, which are not
considered here, include ``colors'' of the broad line emission as
determined by a set of frequency filters.

We define the following three quantities, essentially the
mean, standard deviation, and skewness of a line profile $f$ taken as
a function of $x \equiv \nu/\nu_e$:
\begin{eqnarray}
c_1 & = & \int dx \, x f(x)  \ , \\
c_2 & = & \left[\int dx \, (x - c_1)^2 f(x) \right]^{1/2} \ , \\
c_3 & = & \frac{1}{c_2^{3}} \int dx \, (x - c_1)^3 f(x) \ ;
\end{eqnarray}
here, $f$ is assumed to be normalized so that it integrates to unity.
These observables form a 3D ``moment space'' into which we may map both 
observational data and model predictions in terms of moment vectors
$\vec{c} = (c_1,c_2,c_3)$.

For a given plausible range of model parameters, such as disk radii
and emissivity index $\alpha$, there exists a region $\cal R$ in the
moment space which contains only points that can be generated from the
model line profiles. If a data point falls inside of this region
then the disk paradigm may provide a good general description of the
physics of the observed system.  Conversely, if a data point falls
outside of this surface, then the disk models cannot adequately
account for the observed line profile.

Within the region $\cal R$ there exists a subregion, ${\cal R}_S$,
that consists of all line profile moment vectors from Schwarzschild
systems.  The detection of frame dragging then may amount to observing
a moment vector within $\cal R$ but outside of ${\cal R}_S$.  Of
course, for this scheme to be of practical value, the moment vectors
must respond sensitively to the presence of orbiting disk material
below 6\rg\ in such a way that they are excluded from the region
${\cal R}_S$.  In other words, the Schwarzschild region ${\cal R}_S$
must not fill a large volume of $\cal R$.  Indeed, we have found
that this criterion is met.  Figure~5 illustrates that for a broad
range of disk model parameters (inclination angle of 5$^\circ$ to
40$^\circ$, $0 \leq \alpha \leq 4$, and disk radii less than 15\rg),
Kerr systems with inner disk radii near the innermost stable orbit
generate moment vectors which lie entirely outside of ${\cal R}_S$.

\begin{figure}[tbhp]
 \centerline{
  \epsfxsize=3in\epsfbox{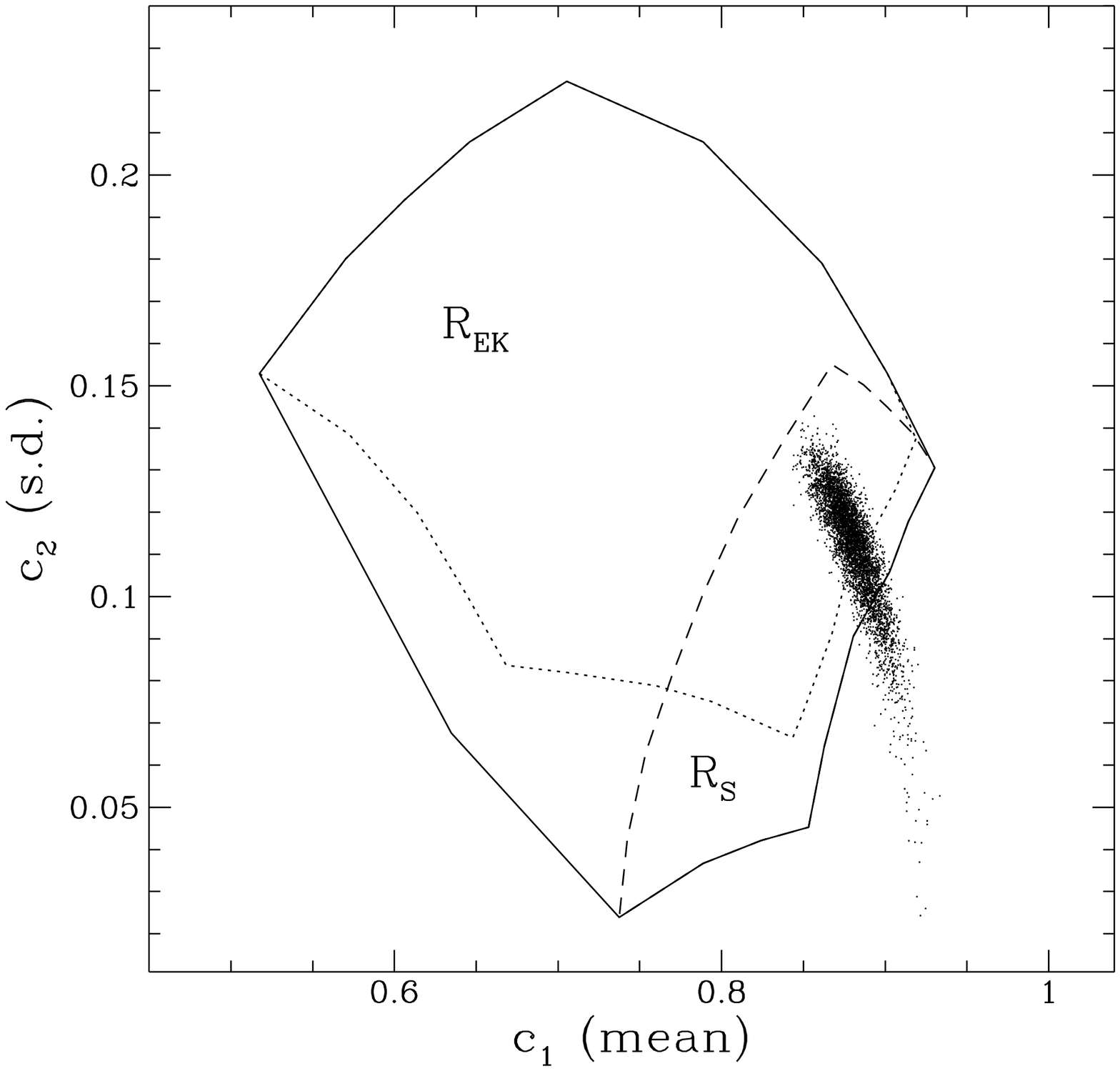}
  \epsfxsize=3in\epsfbox{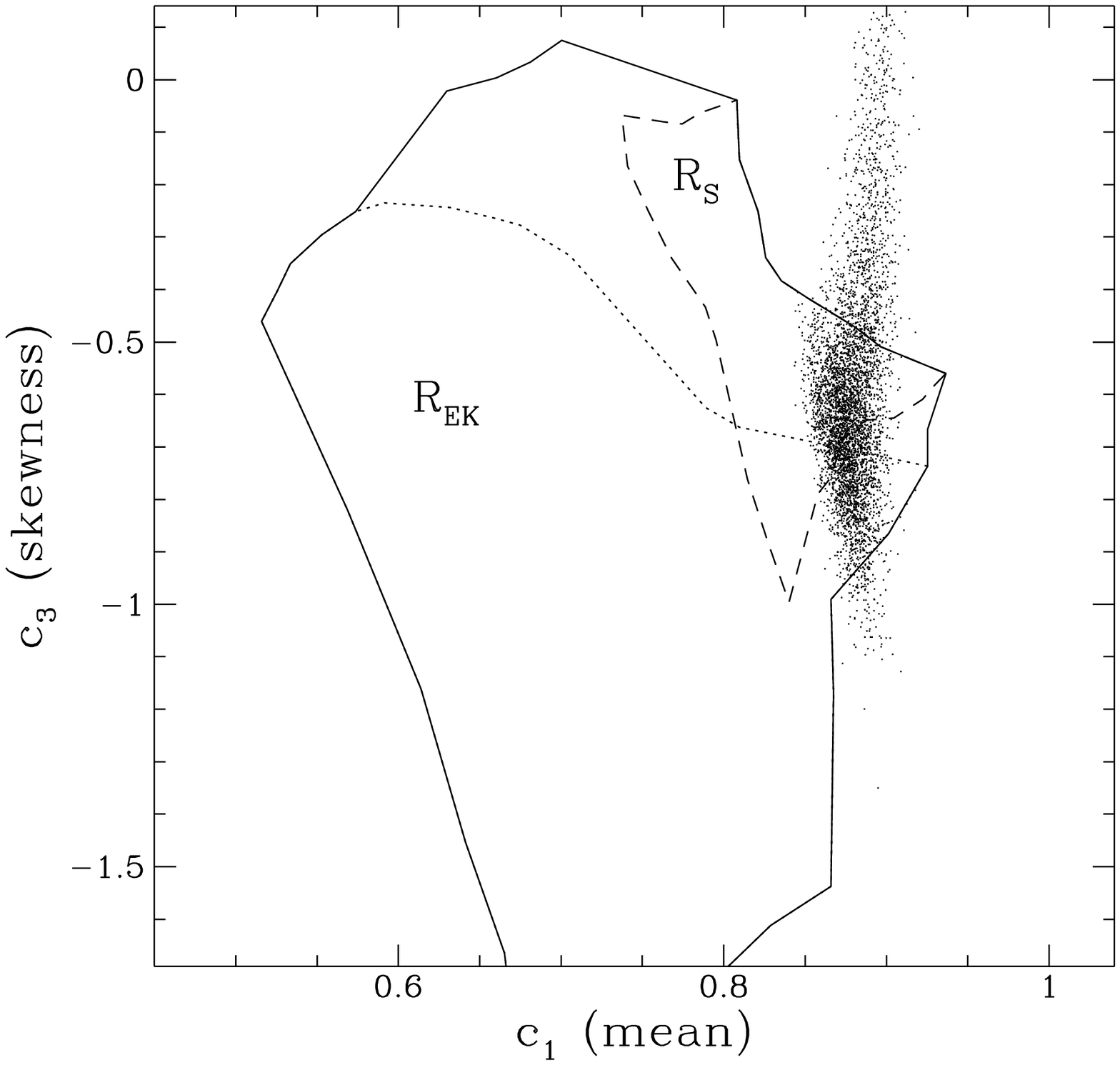}
 }
 \centerline{
  \epsfxsize=3in\epsfbox{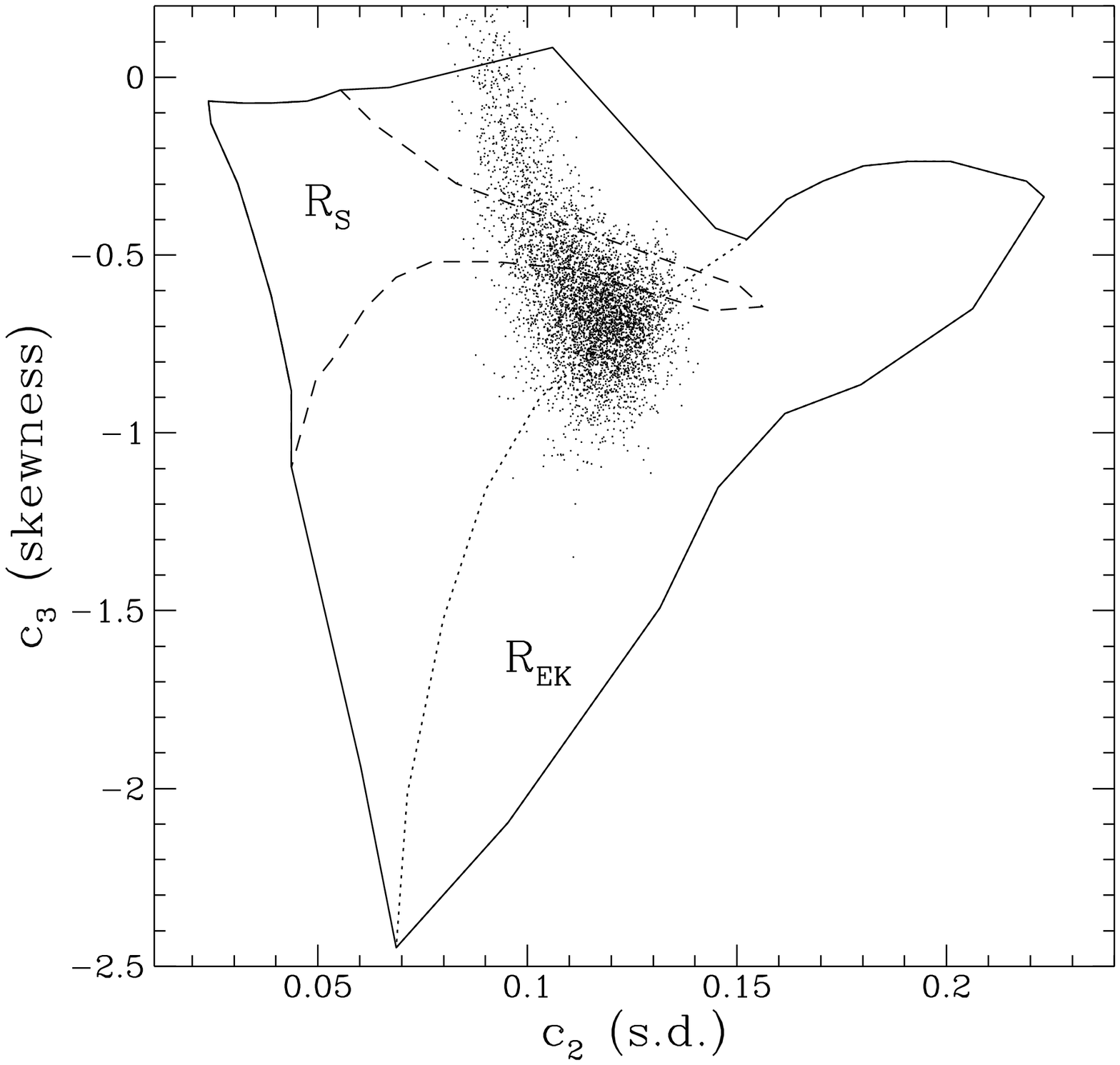}
 }
\caption{
Projections of the 3D line profile moment space defined by the mean,
the standard deviation, and the skewness.  The region inside the solid
black curve contains the predicted vectors from all disk models with
$-4 \leq \alpha \leq 0$, $5^\circ \leq \theta \leq 40^\circ$ and radii
between the innermost stable orbit and 15\rg. The subregions labeled
$R_S$ and $R_{EK}$ correspond to Schwarzschild and extreme Kerr
systems respectively. There is no true overlap between these two
subregions; the apparent overlap is in projection only.  The scatter
points are Monte Carlo realizations of the ASCA \mcg\ iron line data
using the uncertainties given by Tanaka et al. (1995).
}
\end{figure}

This moment-mapping scheme offers the possibility of making fairly
broad statistical inferences from a comparison of models and real
data.  As an example, we again consider the data of Tanaka et al.
(1995).  The line profile data with the published uncertainties were
used to generate Monte Carlo realizations of moment vectors, shown in
Figure~5.  We assumed errors are normally distributed.  The overlap
between these vectors and the regions which correspond to the models
can be used as an indicator of confidence in the models, assuming all
models are equally likely to describe the physics of the system.  The
region $\cal R$ contains $\gsim 60$\% of the Monte Carlo vectors, and
we take this to mean that the relativistic thin disk models as a class
are excluded at a [low] confidence level of $\sim 40$\% on the basis
of the ASCA data. [We are not trying to rule out individual models
here; instead we are favoring or disfavoring a class of models, in
comparison with, say, other unspecified classes that lie outside of
$\cal R$.]

We carry this line of reasoning further and look at the overlap
between the ASCA data and subregions of $\cal R$.  The Schwarzschild
region ${\cal R}_S$ contains $\sim$15\% of the total Monte Carlo
vectors, while a region corresponding to the extreme Kerr systems with
inner radii of 1.25\rg\ contains fewer than 10\%.  Thus the ASCA data
are more likely to be consistent with a Schwarzschild system than this
extreme Kerr system. However, the data also suggest that there is at
least {\em some} disk material below 6\rg; over 40\% of the total
Monte Carlo vectors lie in the region that corresponds exclusively to
models with a rotating black hole, about three times more than in the
Schwarzschild models.  We defer a rigorous statistical analysis
(explicitly considering priors in a Bayesian framework) to another
paper; our strong result here is that frame dragging may be identified
with good confidence after only modest increase in the signal-to-noise
ratio of the data.

\section{Lightcurve of a Line-Emitting Hotspot}  

The integrated line profiles considered in the previous section apply
when optically thick material covers the whole disk, occupying all
possible circular orbits between the inner and outer disk radii. We
now consider effects from localized emitters in the form of bright,
line-emitting hotspots in orbit around a black hole. Such objects
produce a time-varying signal that can reveal valuable information
about the environment of a black hole.  This type of scenario has been
discussed extensively in the literature; recent work includes Bao
(1992), Karas, Vokrouhlick\'{y} \& Polnarev (1992) and Zakharov (1994).

Observations of AGNs at both X-ray and optical frequencies indicate
time variability.  However, at present only optical measurements have
a sufficient signal-to-noise ratio to show any detailed variation in
line profiles.  It is interesting to note that the line profile of
\arp\ which is well-fit by the relativistic disk model has not varied
much in the last two decades.  On the other hand, the line profile (at
least the double peaked component) of many disk emitters have been
seen to vary significantly on time scales as short as a few months.
The origin of this variation is not yet understood. Models with an
inhomogeneous disk (Zheng et al. 1991; Chakrabarti \& Wiita 1994) or
eccentric disk (Eracleous et al. 1995) have been proposed. The hotspot
solutions we discuss here may be used as Green's functions to
construct lightcurves for these more complicated models. For the
present purposes we consider the information content of just one spot.

The variation in line flux from a hotspot arises from at least two
fundamental processes. The first is the frequency shifting of photons
from the Doppler effect as the hotspot orbits the black hole. The
second is flux amplification from gravitational lensing of photons by
the black hole.  Calculation of these two effects, along with the
flight time of photons to a distant observer, produces the lightcurve.
For realistic modeling, other processes may be included, such as
occlusion of photon trajectories by an optically thick accretion disk
or by the surface of the black hole, limb darkening, and tidal
shredding of the orbiting hotspot.  Here, we consider only occlusion
by a disk in the sense that we map only those photons which are
emitted from the side of the hotspot nearest the observer.

In our numerical code for producing accretion disk images we
calculated the frequency shift for each pixel as in Figure~1.  The
gravitational amplification factor is proportional to the solid angle
that a finite patch of emitting material subtends in the sky plane of
the observer.  In the image of an accretion disk, the lensing effect
is calculated implicitly by the mapping of photon trajectories from
the disk to the observer's sky plane.  Lensing is manifested in the
distortion of the disk relative to an image in flat space; regions of
high amplification simply occupy more pixels and thus appear magnified
in the sense of geometric optics. Here, a hotspot is taken to be
arbitrarily small and is therefore unresolvable in an image. The
amplification factor for a given point on the disk is thus evaluated
by mapping adjacent pixels to the disk surface, a procedure which
works quite well given the resolution of our images.

Figures~6~\&~7 show frequency shifts and lightcurves of a hotspot
around a black hole for different disk parameters. The lightcurves in
Figure~6 are similar to those shown in previous work (e.g., Karas,
Vokrouhlick\'{y} \& Polnarev 1992) except that here Kerr and
Schwarzschild cases are directly compared.

\begin{figure}[tbhp]
 \centerline{
  \epsfxsize=3.5in\epsfbox{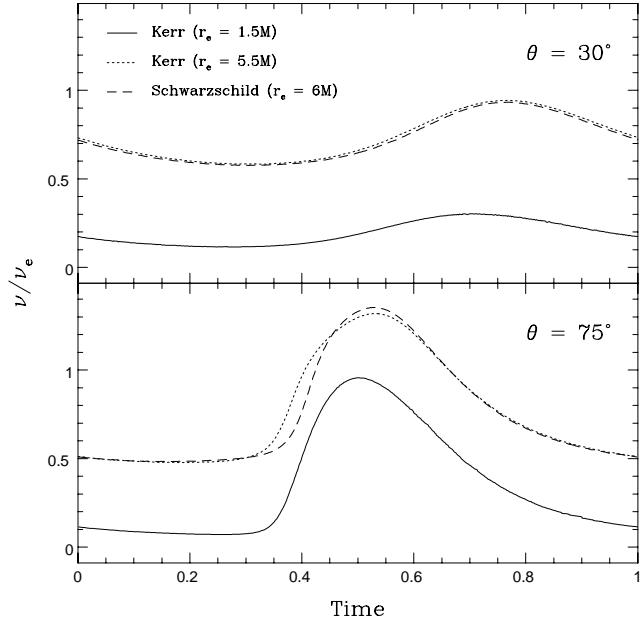}
 }
\caption{
The frequency shift of line emission from a corotating hotspot
in Kerr and Schwarzschild metrics for two different viewing angles $\theta$. 
The radius $r_e$ of the orbit is given for each case
}
\end{figure}

\begin{figure}[bthp]
 \centerline{
  \epsfxsize=3.5in\epsfbox{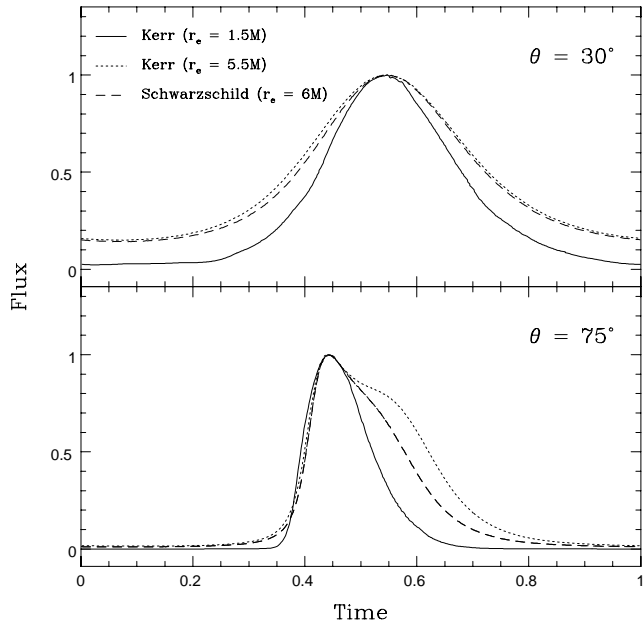}
 }
\caption{
The lightcurve from a line-emitting hotspot in orbit at radius $r_e$
about Kerr and Schwarzschild black holes at two different viewing
angles $\theta$.
}
\end{figure}

In Figure~6 the difference in frequency shifts between a hotspot in a
Kerr metric at a radius $r_e = 5.5$\rg\ and one in orbit at 6\rg\ about
a Schwarzschild black hole is seen to be small for low viewing
angles. However, stable orbits in a Kerr metric below 5.5\rg\ do show
distinguishing features. The most obvious, and perhaps the most
important, is the high-degree of redshifting due to the potential well
of the black hole at small $r_e$.  The profile of the frequency shift
for a hotspot around a Kerr black hole also has a unique form which may be
fit to real data to determine orbital parameters. At high inclination
angle the frequency shift profile alone can serve to distinguish the
extreme Kerr and Schwarzschild geometries, even at radii near $\sim 6$\rg\ 
if the data are of high quality.

Figure 7 contains lightcurves from hotspots in identical
configurations as in Figure 6.  Again, differences between the Kerr
and Schwarzschild systems are not strongly evident at small
inclination angle $\theta$ and large radius ($r_e \gsim 6$\rg), but
are enhanced with increasing $\theta$ and decreasing $r_e$.  Without
any highly distinctive features such as multiple peaks, perhaps these
differences can be exploited only if the angle of inclination can be
determined independently.

It is worth pointing out that the hotspot may also be emitting
continuum radiation.  For example, if the spectrum in the local frame
is a power law $I_c(\nu) \sim \nu^{-\gamma}$, then the time-varying
continuum signal $I_\nu$ from a spot that subtends a solid angle
$\Delta\Omega$ on the observer's sky yields a flux of
\begin{eqnarray}
 F(t) & = & \Delta\Omega(t) \int_{\nu_1}^{\nu_2} \, d\nu I_\nu(t)
 \nonumber 
\\
   & = & \Delta\Omega(t) 
         \int_{\nu_1}^{\nu_2} \, d\nu \left(\frac{\nu}{\nu_e}\right)^3
           I_c(\nu_e)
\label{eq:ftc}
\\
   & = & C \Delta\Omega(t) [{\cal D}(t)]^{3-\gamma} \ ,
%             \int_{\nu_1}^{\nu_2} \, d\nu \nu^{-\beta}\ ,
 \nonumber 
\end{eqnarray}
where ${\cal D} \equiv \nu/\nu_e$ is the redshift which is the same
for all photons from the hotspot at a given time $t$; $\nu_1$ and
$\nu_2$ give the detector bandwidth and $\nu_e$ is the photon
frequency in the source frame.  The constant of proportionality $C$ in
the last line of equations~(\ref{eq:ftc}) depends only on the
detector's bandwidth and the normalization of $I_c$.  With
$\Delta\Omega(t)$ containing the information about amplification from
gravitational lensing and $\cal D$ containing the variations in 
frequency shift (e.g., Fig.~6), it is straightforward to interpret
lightcurves in this case. Interestingly, a continuum spectrum index
of $\gamma = -1$ gives a flux variation of $\Delta\Omega{\cal D}^4$
which is identical to the integrated flux from line emission (e.g. Fig. 6).

\section{Discussion}  

Considerable evidence has accumulated which argues for the existence of
massive compact objects with mass more than $3M_\odot$ --- the
astrophysical definition of black holes.  In AGNs, the observed rapid
variability of enormous luminosity argues strongly for the presence of
a supermassive ($\gsim 10^6 M_\odot$) black hole. The broad optical
emission lines whose profiles are consistent with the relativistic
disk model provide the best kinematic evidence of the existence of a
black hole and also of an associated accretion disk within its deep
potential.  For these reasons, the existence of supermassive black
holes is not disputed.  Therefore, we have focused here on the
information that can be extracted from the local environment of black
holes in AGNs in order to better understand the physics of these
objects.  We hope that in the near future, our effort (development of
the ray tracing code in Kerr metric) may be used to examine real
astrophysical systems and to confirm the frame dragging effect.

This work was largely motivated by the detection of broad Fe lines
from several Seyfert galaxies (Mushotzky et al. 1995).  The fit of the
well-resolved, broad Fe K$\alpha$ line from \mcg\ by the relativistic
disk model is extremely promising (Tanaka et al. 1995).  While we
consider the presence of an accretion disk in this object to be the
best explanation for the observations, there are the following issues
to consider.  The observed equivalent width, around 200 - 400 eV is a
factor of 2--3 higher than the predictions of the simple disk
reflection model with cosmic abundance. Furthermore, in some disk
models, the part of the accretion disk near the black hole would be
too hot due to shock heating to emit the Fe K$\alpha$ line
(Chakrabarti \& Molteni 1995). The equivalent width discrepancy may be
explained by elemental abundances in the disk which can certainly
differ from their cosmic values. The presence of shock heating which
would eliminate the Fe line altogether is best treated
phenomenologically, as follows.

We note that all observations of broad Fe lines in AGNs occur in
Seyfert I galaxies which normally have continuum spectra peaked in the
UV. The flux in the thermal-like UV bump is about 10-20\% of the total
flux. If the UV bump is from thermal emission of accretion disk's
surface, then the thin cool accretion disk has to extend down to
within roughly 10 Schwarzschild radii of the black hole.  At this
radius, hard X-radiation from an external source near the black hole
can produce flourescence lines.  On the other hand, the lack of UV
bump from broad line radio galaxy 3C390.3 is accompanied by a lack of
an extremely broad Fe line (Eracleous et al. 1996).  The comparable
width of the X-ray Fe line and the optical lines supports the
hypothesis that the inner part of the accretion disk is in the form of
a hot torus (Chen \& Halpern 1989).  Therefore, it may be no surprise
that broad Fe lines are only seen in Seyfert I galaxies which have
both a UV bump and a significant hard X-ray flux.

For these reasons we feel justified in choosing diagnostics of a
line-emitting accretion disk as a probe for frame dragging effects in
the nuclei of active galaxies.  As noted before (Laor 1991), perhaps
the best hope of detecting the frame dragging effect is to spot a
stable Keplerian accretion disk at radii below the limit of 6\rg
imposed in a Schwarzschild metric.  Here we made an explicit
comparison of three observable quantities, the integrated line profile
from an accretion disk, the frequency shift of an emission line from a
hotspot, and the lightcurve of hotspot emission. In all cases,
differences between the Kerr and Schwarzschild system were evident.
However, at low inclination angle ($\lsim 30$\deg), the differences
tended to be small. This situation is unfortunate in that X-ray line
emission is more likely to be measured in a system of low viewing
angle, since then the inner part of a galactic nucleus is more likely
to be visible to the observer.

Of the scenarios we discussed above in \S3 and \S4, the most
promising for detecting a signature of frame dragging at low viewing
angle is one in which an integrated line profile can be measured from
a thin disk with emissivity falling off with radius as a power law;
Figure~2 shows the case where the power law index is $-2$.  There is
evidence from optical observations that such an emissivity function is
realistic (e.g., Chen \& Halpern 1989). The evidence from X-ray lines
is less compelling primarily because the available data are noise
limited, but such an emissivity function is nonetheless consistent with
observations (e.g., Tanaka et al. 1995). In this case, the presence of
material in stable orbits below $\sim 6$\rg\ becomes quite clear; in
Figure~2, for example, the net flux enhancement due to material
between 1.25\rg\ and 6\rg\ is roughly a factor of 1.5.  This enhancement
is close to the amount needed to eliminate the discrepancy between the
observed equivalent width in the Fe line of \mcg\ and the prediction
of a thin disk model.

In constructing Figures~2--4 we chose parameters which were similar to
those deduced from \mcg\ by Tanaka et al. (1995). In their paper
Tanaka et al. showed that an extreme Kerr disk and a Schwarzschild
system give very similar $\chi^2$ values, with an inner disk radius of
about 8\rg. In light of the present work, this is not altogether
surprising, as we find the differences between the Kerr and
Schwarzschild metric are generally not strong beyond 6\rg.  
The X-ray data from \mcg\ are noisy, and evidently the $\chi^2$
surface for the model fitting is fairly flat; the uncertainty in the
inner disk radius appears to be $\sim$1.5\rg\ at the 1-$\sigma$ level.
Thus, future observations of this object with increased resolution
hold the promise of detecting stable orbits within 6\rg.

While line profiles should be fit explicitly to determine model
parameters, we have demonstrated (\S3.1) that the three lowest moments
of the profiles can provide a discriminating measure of confidence in
broad classes of disk models. Specifically, the mean, standard
deviation, and skewness of a normalized profile defines a point in a
3D moment space and the position of points in this space is sensitive
to emission from material below 6\rg.  Figure 5 is demonstration that
maps in this space may be used effectively to identify frame dragging
in a broad context of disk models.  Formally, the ASCA data suggest
that black hole rotation is roughly three times more likely to be
finite than zero.

We must caution that the use of the ASCA data here is for the purposes
of example; ideally, more data than were published by Tanaka et al.
(1995) should be used to fit the continuum background. Furthermore,
Iwasawa et al. (1996) have reported significant temporal variations in
the line profile on scales shorter than the 4 day integration time.
Nonetheless, these issues do not greatly weaken our conclusions
regarding the possibility of detecting a signature of frame dragging
in line profiles.

An additional caveat is that we have dealt only with simple idealized
systems consisting of a geometrically thin accretion disk emitting in
a single atomic line.  The thin-disk approximation is certainly good
for an accretion disk at a few hundred Schwarzschild radii where many
optical lines are produced.  However, for the accretion disk at a few
\rg, the ratio of disk thickness relative to the radius at a given
point may not be negligible. Furthermore, non-Keplerian motion in the
line-emitting material may modify the line profile to a certain
degree.  Electron scattering in the inner part of accretion disk can
also change the energy of line-emission photons and hence can broaden
the line profile at local emitting frame.  However, we do not expect
that the line would be distorted beyond recognition by these
uncertainties, but there may be further challenges in identifying a
signature of the frame dragging effect.

We have presented results from a general purpose null-ray tracing code
applied to the problem of emission from an accretion disk.  Clearly,
we can advance the level of modeling beyond the thin disk
approximation.  Thus as the resolution of observed X-ray line emission
increases in the near future, we will be able to model more
sophisticated physics to probe the environments of rotating black
holes.

\subsection*{Acknowledgements}

We thank Stirling Colgate, Daniel Holz and Paul Wiita for helpful
discussions and comments.  The criticism and suggestions by the
anonymous referee, particularly with regard to quantitative measures
of line profiles, are greatly appreciated.  We are endebted to John
Archibald Wheeler for providing specific improvements to the
manuscript and especially for guiding us in the search for
astrophysical footprints of gravitomagnetism.  This work was supported
by US Department of Energy through the LDRD/IP program and by the NASA
HPCC program. BCB acknowledges partial support from NSF grant
PHY-95-07695. The Cray Supercomputer used in this investigation was
provided through funding from the NASA Offices of Space Sciences,
Aeronautics, and Mission to Planet Earth.

% \newpage

\def\ref{\bibitem{}}

\end{document}